# A Scenario Based Approach For Dealing With Challenges In A Pervasive Computing Environment

Divyajyothi M G[1], Rachappa[2] and Dr. D H Rao[3]

[1,2]Research Scholar, Department of Computer Science, Jain University, Bangalore
[3]Principal and Director, Jain College of Engineering, Belgaum

**ABSTRACT**

*With the surge in modern research focus towards Pervasive Computing, lot of techniques and challenges needs to be addressed so as to effectively create smart spaces and achieve miniaturization. In the process of scaling down to compact devices, the real things to ponder upon are the Information Retrieval challenges. In this work, we discuss the aspects of multimedia which makes information access challenging. An Example Pattern Recognition scenario is presented and the mathematical techniques that can be used to model uncertainty are also presented for developing a system that can sense, compute and communicate in a way that can make human life easy with smart objects assisting from around his surroundings.*

## 1. INTRODUCTION

Tremendous growth and research contributions are going on towards Mark Weiser's vision [4][5] on developing a system that can sense, compute and communicate in a way that can make human life easy with smart objects assisting from around his surroundings. In this process of migrating towards a smart environment, the real challenges to ponder upon are the performance issues, data management, software maintenance, energy efficiency, trust, security and privacy of the computing device to be designed[1][2][3][6][7]. The Demand for pervasive multimedia services has widely increased with Ubiquitous computing being expected in almost all areas of health care, entertainment, digital libraries, hotels, class rooms, Smart campuses, automobiles, streets, airports, social networks. With the amount of multimedia data we have, data analysis, indexing, retrieval, distribution, and management becomes even more challenging subject to the fact of adopting all these requirements that suits the human being mode of thinking, expectations and vision [9].This is where context aware multimedia computing becomes extremely important. In this paper, the aspects that make the multimedia retrieval challenging is discussed. An example scenario is presented with explanation on tools that can be readily used to model uncertainty are described.

## 2. MULTIMEDIA RETRIEVAL CHALLENGES

### 2.1. Real –Time Constraints

Multimedia involves very large amounts of data. Multimedia retrieval refers to extracting semantic information from this large amount of data available in various forms. Thus multimedia retrieval needs efficient techniques and algorithms for redundancy elimination, feature extraction and categorization.

## 2.2. Bridging the Semantic Gap

There is a semantic gap between the semantics the searcher attaches to the visual or any multimedia information and the semantics which is extracted from the digital information stored for the multimedia objects. The semantic content needs to be structured and summarized and therefore is a daunting task.

## 2.3. "Multi" in Multimedia

Analysing characteristics of all the possible myriad forms of media and retrieving it in a standard way from these vast amounts of multimedia information for all of them is quite a daunting task. Some forms of multimedia: Text, Sketches, Videos, Graphical Images, Speech, Sound, Movies.

## 2.4. Effective Multimedia Search Engines

Need an effective search mechanism while answering a query to a document database. The most common problems, which may occur in this process, are synonymy and polysemy. Synonymy – Search Engine detecting any subject S that may not exactly present in any article A. Polysemy -- some words may have many meanings.

## 2.5. Protocols for Multimedia Networks

For Instance, in order to constantly receive unobtrusive connectivity and response from network devices embedded in the environment, the computing speed should be invariably good and this can be made possible if parallel computation can take place.

## 2.6. Multimedia languages for Multi-Channel Content

We need more possible ways to distribute our content video, audio, Speech, images, Text, for excellent ubiquitous behaviour. Efficient scalable codecs needs to be proposed for effective universal multimedia access.

## 2.7. Multimedia Services for Intelligent Pervasive Computing

It's necessary to address the security and privacy issues involved while performing secure transactions in a ubiquitous environment. More hybrid security measures have to be implemented keeping in mind the performance of devices that are used in pervasive computing environments.

## 2.8. Scalable Algorithms / Techniques

It's time we think about whether the existing retrieval, indexing, mining, streaming, delivery, personalization algorithms are easily feasible with compact pervasive devices. If not new algorithms have to be formulated based on the existing foundations we have. In such cases quantum mechanics techniques may provide useful results.

## 3. EXAMPLE SCENARIO

### 3.1. Complete Pattern Recognition System

#### 3.1.1 Sensor for Feature Extraction

Requires a sensor for Feature Extraction mechanism, requires a Classification / Description Scheme which is based on learning strategy/paradigms.

   a. Supervised Learning – based on availability of a training set (set of patterns).
   b. Unsupervised Learning – based on statistical regularities of patterns Decision Theoretic.
   c.

#### 3.1.2 Classification Scheme

Requires a Classification / Description Scheme which uses one of the following:
   a. Statistical (Decision theoretic) is based on statistical characterizations of patterns which are generated by a probabilistic system. For a Probabilistic system, naive Bayes classifier can be very effectively used because it works on the availability of a particular training set irrespective of other features and its results are based on maximum likelihood property. Example for document classification: Documents can be classified based on many ways by their content, subject, Text and Graphical portions, Images and plain text, mathematical equations and numbers, noun sorters, mass dividers. In general, each document classification has its own classification challenges.

   b. Syntactic (Structural) – based on structural interrelationships of features, should have a clear structure of the patterns.  An appropriate grammar is the core of any type of syntactic pattern recognition process. One must make sure that the Grammars are established from a priori knowledge about the objects or scenes to be recognized. Syntactic Classifiers: Representing structural information in images can be effectively done using Fuzzy set Theory. All imprecise relationships between objects can be defined as spatial fuzzy sets. Formal language can be readily used to represent such structures. Fuzzy grammars can be generated for the objects to be recognized. All differences in the structures of the classes must be encoded as different grammars. Another example can be Diagnosis of the heart using ECG measurements.

   c. Neural Classifiers - based on bionics-related concepts in recognizing patterns. Bionics refers to the science of applying biological concepts to electronic machines. This concept is made use by the neural approach that applies these biological concepts to machines in order to recognize patterns. As an outcome of this effort that field of artificial neural networks has emerged and we can see some interesting results. Neural Classifiers: Neuro Excel Classifier is one of the most efficient, quick, powerful and an easy-to-use neural network software tool that is most widely being used for classifying data in Microsoft Excel. The main objective of this classifier is to aid experts in the design process of real-world data mining and pattern recognition tasks. One of the major benefits of this classifier is its ability to hide the underlying density, Thickness, complexity of neural network processes by providing graphs and statistics for the user so that the results can be easily understood. The algorithms and techniques used in Neuro Excel Classifiers are only those which are reliable and which are proven to be efficient. Another feature is its ability to amalgamate flawlessly with Microsoft Excel. [8]

Apply appropriate Algorithms for the pattern recognition based on the target system.

## 4. MODELLING UNCERTAINTY

Every Situation in the world around us can be represented as a mathematical model. All these models are established using the building blocks of Set Theory. Set theory is a branch of mathematics that deals with the properties of sets. According to classical set theory, the membership of elements in a set is based on a bivalent condition, stating that an element may either belong or may not belong to the set.

Generalization of the classical set theory gives us the fuzzy set theory, in which the membership of elements in a set is described with the aid of a membership function valued in the real unit interval [0, 1]. The method of assessing the membership of elements in classical set theory is based on a crisp condition of whether the element belongs to or does not belong to any given set S. Therefore the entire process is being done in binary terms.

### 4.1. Fuzzy Set Theory

Fuzzy sets are considered to be an extension of classical set theory. Such imprecise concepts, classical set theory fails to handle since it uses the principle of bivalent condition. So when it comes to expert systems and recognition system, Fuzzy set theory serves better as it has the ability of handling effectively the most inherent and imprecise concepts. It is most widely used mathematical method in modern research because it is much more organised and can handle imperfect knowledge and vagueness in an intelligent manner. We should be aware that not all concepts can be converted in the form of an equation. For example, consider the problem of expressing the term "hotness" in the form of a mathematical equation. Since, "hotness" is not a quantity; it cannot be expressed in the form of an equation. But still if one asks common people they have an idea of what is "hot", and agree that there is no sharp cut-off between "hot" and "not hot", where something is "hot" at N degrees but "not hot" at N-1 degrees.

Thus we can define a fuzzy set on a classical set is defined as follows:

$$\tilde{A} = \{(x, \mu_A(x)) \mid x \in X\}$$

In the above equation, the membership function $\mu_A(x)$ enumerates whether the elements x belong to the fundamental set X or not. Based on the bivalent condition, the elements in the set can either take the value of 1 or 0. If the element is seen mapping to the value 0, it means that the element does not belong to the given set X. Likewise if the element is seen mapping to 1, it is fully a member of the set. The elements that range in between are said to be the fuzzy members. Consider a fuzzy set C, where C = {(3,0.3), (4,0.7), (5,1), (6,0.4)}. Using standard fuzzy set theory notation, this would be computed as C = {0.3/3, 0.7/4, 1/5, 0.4/6} From the above notation, it can be observed that any value with a membership grade of zero does not appear in the expression of the set. Finding the membership grade of the fuzzy set C at 6 using the standard notation is $\mu_B(6) = 0.4$.

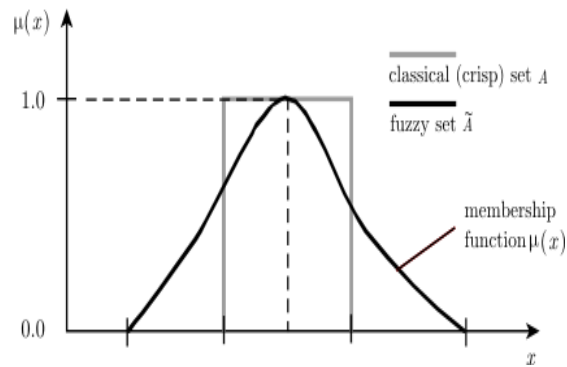

Figure1: Fuzzy Set and Crisp Set

### 4.1.1 Degrees of Truth

The real numbers in the interval [0, 1] are usually used to represent the Degrees of Truth.
If the values are the extreme points (0 and 1), then they represent absolute falsity and absolute truth respectively whereas the values ranging in between the extreme points (0 and 1), represent the intermediate truth degrees.

Hence, when a system of logic uses these degrees of truth it is called fuzzy logic. The logical operations in such systems are typically defined as follows:

$$\neg P = 1 - P$$

$$P \lor Q = \max(P, Q)$$

$$P \land Q = \min(P, Q)$$

As a result, such systems employing degrees of truth allow us to assess sentences involving more than one vague and myriad property such as warm, old, strong, fit, happy, bright, sad, cold, and coldest and so forth. As a result we have a technique to potentially tackle vagueness. [11][12][13]

### 4.1.2 Logic based on Fuzzy Sets

Logic based on the concept of fuzzy sets, in that any membership that is expressed using varying probabilities or degrees of truth with values ranging from 0 (does not occur) to 1 (definitely occurs).

There are a lot of practical uses that Fuzzy logic offers in making a considerable impact in engineering control systems, computing, medicine and healthcare, where pervasive computing is allowing self-care rather than professional care [14].

It is more convenient to programming a series of logical conditions and corresponding actions into a machine. Using fuzzy logic for the same allows the values of the propositions involved to come straight away from the machine's sensors.

Few Examples are the usage of sensors in devices such as thermometers, motion detectors, refrigerators, washing machines.

Commercial applications of Fuzzy Logic came up as early as in the 1990s. Many other products using fuzzy logic include camcorders, microwave ovens, Toasters, Vacuum cleaners, dishwashers. Other applications include expert systems in a pervasive environment using multiple sensors, interactive devices, and computerized speech- and handwriting-recognition programs with Interfaces for modes of interactions between people and pervasive computing devices.

## 5. SET THEORY APPLICATION AREAS

### 5.1. Electronic Commerce

Rough set theory is used nowadays in Electronic Commerce (EC) for data mining. Everyday large volumes of data are collected by the EC sites. This data comprises a lot of secure, valuable information about customers, products, transactions, communication address and so on. This large amount of data can be better used by the site management to extract the unknown knowledge or trends hiding in the data, and to arrange their products according to the buyers' preference and take appropriate selling, security and authentication policies.

### 5.2. Pattern Recognition

Rough set theory is widely used in combination with neural network theory for pattern recognition. Using All Set Theory and Set Pair Analysis (SPA) many new methods of pattern recognition have been proposed. These methods have proven to be much more valid and effective when compared to existing conventional ways of pattern recognition.

### 5.3. Computer Networks

Rough Set theory has also found its wide application in computer network fault diagnosis, anomaly intrusion detection in computer networks. It has served as an excellent tool for dealing with vagueness and imperfections.

### 5.3. Pervasive Computing

Lot of Service Match Making Algorithms for Pervasive Computing are based on Rough Set Theory. Also Rough Set theory is being used for creating a user aware TV program and Settings in a pervasive environment [10].

## 6. CONCLUSION

Most of the time, probability is being confused with Degrees of truth. This should not be the case. Consider the task of flipping a coin. It would be incorrect to say that this task has a 50%-50% chance of being or not being in F. Though flipping a coin definitely results either in a heads or tails, giving it a 50% chance is incorrect. So this random event must be given the value 1 for its degree of truth because irrespective of the situation, one of the sides either heads or tails definitely appears. Degrees of truth should also never be confused with an unknown or varying truth value.

For Example, Consider the sentence-"The month of July is usually a monsoon day in some parts of the world". Though the degree of truth value does not fall on the extreme points 0 or 1, it can be still considered as a definite value. Repeated observations made on the same day do not give us different values. Thus an alternative to probability theory can be the mathematical theory known as possibility theory which can be used to deal with certain types of uncertainty. This theory was first introduced in 1978. Basically this theory was developed as an extension to the fuzzy logic and fuzzy set theory by Prof Zadeh. This theory makes use of the necessity and possibility of any random event whereas the probability theory makes use of only probability to decide on the likeliness of an event to take place.

Choosing the ELECTIC VIEW can be an alternate solution, which agrees / accepts both the interpretations: depending on the situation, one has to select one of the two interpretations for pragmatic, or principled, reasons.

## *7.* ACKNOWLEDGMENTS

We would like to express our gratitude to the person who made completion of this paper possible. We are deeply indebted to our Prof. Dr. D. H. Rao for his constant help, suggestions and motivation to study and work. His tremendous knowledge and stimulating suggestions has helped us a lot in completing our work effectively.

## *8.* REFERENCES

**Authors**

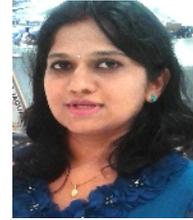

Mrs DivyaJyothi M.G. is currently working as Lecturer at the Department of Information Technology, Al Musanna College of Technology, Sultanate of Oman. Her teaching interests include Pervasive Computing, Firewalls and Internet Security Risks, E-Commerce, Computer Networks, Intrusion detection System, Network Security and Cryptography, Internet Protocols, Client Server Computing, Unix internals, Linux internal, Kernel Programming, Object Oriented Analysis and Design, Programming Languages, Operating Systems, Image Processing, Web Design and Development, etc. Her most recent research focus is in the area of Pervasive Computing. She received her Bachelor and Master Degree in Computer Science from Mangalore University, She bagged First Rank in Master's Degree at Mangalore University. She has been associated as a Lecturer of the Department of Information Technology since 2007. She has worked as Lecturer at ICFAI Tech., Bangalore, T John College for MCA, Bangalore, Alva's Education Foundation Mangalore. She has guided many project thesis for UG/PG level.

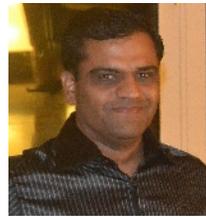

Mr. Rachappa is currently working as Lecturer at the Department of Information Technology, Al Musanna College of Technology, Sultanate of Oman. His teaching interests include Computer Security, Pervasive Computing, E-Commerce, Computer Networks, Intrusion detection System, Network Security and Cryptography, Internet Protocols, Client Server Computing, Unix internals, Linux internal, Kernel Programming, Object Oriented Analysis and Design, Programming Languages, Operating Systems, Web Design and Development, etc. His most recent research focus is in the area of Security Challenges in Pervasive Computing. He received his Bachelor Degree in Computer Science from Gulbarga University, Master of Science Degree from Marathwada University and Master of Technology in Information Technology Degree from Punjabi University (GGSIIT). He has been associated as a Lecturer of the Department of Information Technology since 2006. He has worked as Lecturer at R.V. College of Engineering, Bangalore. He has guided many project thesis for UG/PG level. He is a Life member of CSI, ISTE.

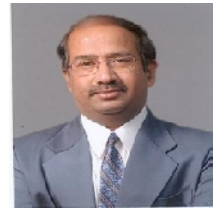

Dr. D H Rao is currently working as a Dean, Faculty of Engineering, VTU, Belgaum. Principal and Director, Jain College of Engineering, Belgaum.He is the Chairman, Board of Studies in E & C Engineering, VTU in Belgaum. He is a Member, Academic Senate in VTU Belgaum. He has over 100+ publications in reputed journals and conferences. He obtained B.E. (in Electronics from B.M.S. College of Engineering), M.E. (from Madras University), M.S. (University of Saskatchewan, Canada) Ph.D. (Univ. of Saskatchewan, Canada).